# AXION STRING CONSTRAINTS

R. A. Battye *and* E. P. S. Shellard

*Department of Applied Mathematics*
*and Theoretical Physics*
*University of Cambridge*
*Silver Street, Cambridge CB3 9EW, U.K.* *

## Abstract

We study the production of cosmological axions in the standard scenario in which a global string network forms at the Peccei-Quinn phase transition. We make detailed calculations of the axions produced by string loops, comparing these with estimates of other contributions from long strings and domain walls. We delineate key uncertainties in string network evolution, the chief of which is the magnitude of radiative backreaction. We conclude that axions produced by these topological contributions provide the significant cosmological constraint on the symmetry breaking scale $f_{\rm a}$ and the axion mass $m_{\rm a}$.

## 1 Introduction

The nature of the dark matter of the universe remains an outstanding question facing cosmology today. Of the possible candidates, the axion is one of the most promising. In axion models an extra global $U(1)_{\rm PQ}$ symmetry is added to the standard particle physics model to solve the strong CP problem [1]. This symmetry is broken at a high energy scale $f_{\rm a}$ and the resulting pseudo-Goldstone boson—the axion—subsequently acquires an extremely small mass through 'soft' instanton effects at the QCD phase transition [2]. Because of substantial redshifting before it becomes massive, the axion is an ideal cold dark matter candidate. Viability, however, is dependent on compatibility between the predicted cosmological axion density and observational and nucleosynthesis constraints on the total and baryonic densities [3]. These constraints provide a significant upper

---

bound on the symmetry breaking scale $f_a$ which is near conflict with astrophysical bounds from red giants and SN1987a, $f_a \gtrsim 10^9$–$10^{10}\,\mathrm{GeV}$ [4].

If the universe passes normally through the Peccei-Quinn phase transition, a network of global strings will form by the Kibble mechanism [5]. (We note here that we will not be considering the alternative string-less scenario in which this transition occurs before an inflationary epoch—see, for example, ref. [6].) The decay of this string network into axions provides the dominant contribution to the overall axion density $\Omega_a$ [7]. However, the dependence of $\Omega_a$ on the energy scale $f_a$ has been the subject of much debate because radically different string radiation mechanisms have been proposed [7,8,9]. In a recent publication [10], we investigated the string radiation spectrum in great detail, using both analytic and numerical techniques. These findings were in broad agreement with the original work of Davis *et al.* [7,9] and contrary to the predictions of Sikivie *et al.* [8]. Taken at face value these results support a string constraint which would rule out the axion in the standard scenario. However, we also noted that, although the underlying physics of the earlier work was correct, the model employed for string evolution was too simplistic.

In this letter we present a model for the evolution of an axion string network based on a marriage of numerical results for local strings [11,12] and the now-validated analytic radiation calculations for global strings which employ the Kalb-Ramond action [13,14,10]. Using this model, we calculate the density of axions radiated by string loops from the scale-invariant network properties (a previously overlooked source [15,16]). Using estimates for these network parameters, the loop contribution to the axion density is compared to previous estimates for the contribution from long strings, domain walls and the homogeneous zero momentum mode.

## 2  The evolution of an axion string network

The evolution of axion strings is qualitatively very similar to the evolution of local strings due to their dynamical correspondence—as demonstrated numerically [10]. The additional long-range Goldstone field acts primarily to renormalize the string tension and energy density,

$$\mu \approx 2\pi f_a^2 \ln(t/\delta),$$

where the string core width is $\delta \sim f_\mathrm{a}^{-1}$ and we assume the typical curvature radius of the strings at a time $t$ is $R \sim t$. Quantitatively on small scales $\ell \ll t$, global strings are affected by enhanced radiation backreaction; typically in a cosmological context axion radiation will be three orders of magnitude stronger than the weak gravitational radiation produced by local strings. This difference will alter small-scale features such as string wiggliness and loop creation sizes, but not the more robust large-scale network properties. As observed in numerical simulations of local strings [11,12], large-scale properties are remarkably independent of small-scale effects. As a first approximation, therefore, we employ parameter values found by these simulations.

After formation, global strings experience a significant damping force due to the relatively high radiation background density. This frictional force eventually becomes negligible and the strings begin to oscillate relativistically and to radiate axions from the time $t_* \sim 10^{-20}(f_\mathrm{a}/10^{12}\,\mathrm{GeV})^{-4}\,\mathrm{sec}$.

From $t_*$ onwards, we can expect the axion strings to approach a scale-invariant regime in which the network properties remain constant relative to the horizon size $d_H = 2t$. The overall density of strings splits neatly into two distinct parts, the long strings with length $\ell > t$ and small closed loops with $\ell < t$. The long strings have an overall energy density

$$\rho_\infty = \zeta \frac{\mu}{t^2}\,, \qquad (1)$$

where the correlation length scale of the Brownian network is $\xi \approx \zeta^{-1/2} t$ where $\zeta \approx 13$.

To describe scale-invariant loop creation and decay we must define several further parameters: First, we employ $\alpha$ to characterize the average loop creation size, that is, $\langle \ell \rangle = \alpha t$. Secondly, the loop backreaction parameter $\kappa$ describes the radiation power from loops which is given by [13]

$$P = \Gamma_\mathrm{a} f_\mathrm{a}^2 = \kappa \mu\,, \qquad (2)$$

where $\Gamma_\mathrm{a}$ is a factor dependent on the loop trajectory, but not its size, which is estimated to be $\langle \Gamma_\mathrm{a} \rangle \approx 65$ [13,17] (exploiting similarities with gravitational radiation). As the loop decays into axions, its radius shrinks linearly

$$\ell = \ell_\mathrm{i} - \kappa(t - t_\mathrm{i})\,, \qquad (3)$$

where $t_i$ is the loop creation time and $\ell_i = \ell(t_i)$. Typically, in a cosmological context we have $\kappa \approx (\Gamma_a/2\pi)[\ln(t/\delta)]^{-1} \approx 0.15$. Finally, we define the long string backreaction parameter $\gamma$; lengthscales below $\gamma t$ are smoothed by radiative damping in one Hubble time. Naively, one might expect $\gamma \approx \kappa$, but the study in ref. [10] indicated that long string radiation, primarily in the second fundamental mode, was somewhat weaker with $\gamma \sim 0.1\kappa$. The significance of $\gamma$ is that it should set the minimum loop creation size, that is, we expect $\gamma \lesssim \alpha \lesssim \kappa$. Given these assumptions and energy conservation considerations, our scale-invariant model implies that the number density of loops in the interval $\ell$ to $\ell + d\ell$ is given by

$$n(\ell,t)d\ell = \frac{\nu d\ell}{t^{3/2}(\ell + \kappa t)^{5/2}}, \qquad \ell \leq \alpha t, \tag{4}$$

where $\nu \approx 0.40\zeta\alpha^{1/2}$.

Near the QCD phase transition the axion acquires a mass and network evolution alters dramatically because domain walls form [18], with each string becoming attached to a wall [5]. Initially, the mass is temperature-dependent, but it only becomes significant when the Compton wavelength falls inside the horizon, that is, $m(\tilde{t})\tilde{t} \sim 0.75$ at the time

$$\tilde{t} \sim 8.8 \times 10^{-7} \Delta^2 \left(\frac{f_a}{10^{12}\,\text{GeV}}\right)^{0.36} \left(\frac{\bar{m}_a}{6 \times 10^{-6}\,\text{eV}}\right)^{-2} \left(\frac{\mathcal{N}_{\text{QCD}}}{60}\right)^{0.5} \text{sec}, \tag{5}$$

where $\Delta$ is a constant of order unity which quantifies parameter uncertainties at the QCD phase transition*,

$$\Delta = \left(\frac{\bar{m}_a}{6 \times 10^{-6}\,\text{eV}}\right)^{0.82} \left(\frac{\Lambda_{\text{QCD}}}{200\,\text{MeV}}\right)^{-0.65} \left(\frac{\mathcal{N}_{\text{QCD}}}{60}\right)^{-0.42}, \tag{6}$$

with the final axion mass $m_a = \bar{m}_a(f_a/10^{12}\,\text{GeV})^{-1}$ and the mass at temperature $T$ given by $m(T) = 0.1 m_a (\Lambda_{\text{QCD}}/T)^{3.7}$ [19].

Large field variations due to the strings collapse into domain walls at $\tilde{t}$. Subsequently, these domain walls begin to dominate over the string dynamics when the force from the surface tension becomes comparable to the tensional force due to the typical string curvature $\sigma \sim \mu/t$,

$$t_w \sim 1.7 \times 10^{-6} \Delta^2 \left(\frac{f_a}{10^{12}\,\text{GeV}}\right)^{0.36} \left(\frac{\bar{m}_a}{6 \times 10^{-6}\,\text{eV}}\right)^{-2} \left(\frac{\mathcal{N}_{\text{QCD}}}{60}\right)^{0.5} \text{sec}. \tag{7}$$

---

* Note that to calculate $\tilde{t}$, we assume an effective number of massless degrees of freedom $\mathcal{N}$ in an epoch when its actual value is falling.

The demise of the hybrid string–wall network proceeds rapidly [5], as demonstrated numerically [15,20]. The strings frequently intersect and intercommute with the walls, effectively 'slicing up' the network into small oscillating walls bounded by string loops. Multiple self-intersections will reduce these pieces in size until the strings dominate the dynamics again and decay continues through axion emission.

## 3 Calculation of the loop contribution to the axion density

Given the loop distribution (4), we can calculate the energy density of emitted axions. The radiation spectrum from a loop of length $\ell$, averaged over various loop configurations, is given by

$$\frac{dP_\ell(\omega)}{d\omega} = f_a^2 \ell\, g(\ell\omega), \qquad (8)$$

where the function $g(x)$ is normalised by

$$\int_0^\infty g(x)dx = \Gamma_a, \qquad (9)$$

and $\Gamma_a$ is defined in (2) (this approximates the loop spectrum which is actually discrete). We shall assume that loops are at rest, because any initial velocity will be redshifted and the net error when averaged isotropically over all loops should be relatively small.

The energy density of massless axions emitted at time $t_1$ in an interval $dt_1$ with frequencies from $\omega_1$ to $\omega_1 + d\omega_1$ is

$$d\rho_a(t_1) = dt_1 d\omega_1 f_a^2 \int_0^\infty d\ell\, n(\ell, t_1) \ell\, g(\ell\omega). \qquad (10)$$

Assuming $\mathcal{N}$ constant, the spectral density can be calculated by integrating over $t_1 < t$, taking into account the redshifting of both the frequency, $\omega = a(t_1)/a(t)\omega_1$, and the energy density, $\rho_a \propto a^{-4}$. Neglecting the slow logarithmic dependence of the backreaction scale $\kappa$, we have

$$\frac{d\rho_a}{d\omega}(t) = f_a^2 \int_{t_*}^t dt_1 \left(\frac{t_1}{t}\right)^{3/2} \int_0^{\alpha t_1} d\ell\, \frac{\nu\ell}{(l+\kappa t_1)^{5/2}}\, g\left[(t/t_1)^{1/2}\omega\ell\right]. \qquad (11)$$

Under the substitution $x = \ell/t_1$, $z = \omega x(tt_1)^{1/2}$, the range of integration is transformed and (11) becomes [16]

$$\frac{d\rho_a}{d\omega}(t) = \frac{4f_a^2 \nu}{3\omega\kappa^{3/2}t^2} \int_0^{\alpha\omega t} dz\, g(z) \left[\left(1 + \left(\frac{z}{\omega\kappa t}\right)\right)^{-3/2} - \left(1 + \frac{\alpha}{\kappa}\right)^{-3/2}\right], \qquad (12)$$

since the contribution from the lower limit can be shown to be zero for the range of $\omega$ under consideration. This implies that the peak contribution to the axion density comes from those axions emitted just before wall domination.

One can approximate the integrals of $g(z)$ by noting that the dominant contribution comes in the range $4\pi < z < 4\pi n_*$, where $n_*$ is the mode beyond which the radiation spectrum of loops can be truncated due to backreaction. Assuming $4\pi n_* << \omega \kappa t$ and using the normalisation condition (9), the integral (12) becomes

$$\frac{d\rho_a}{d\omega}(t) \approx \frac{4\Gamma_a f_a^2 \nu}{3\omega \kappa^{3/2} t^2} \left[1 - \left(1 + \frac{\alpha}{\kappa}\right)^{-3/2}\right]. \qquad (13)$$

From this expression we can obtain the spectral number density of axions $dn_a/d\omega = \omega^{-1} d\rho_a/d\omega$. Integrating and comparing with the entropy density of the universe, $s = 2\pi^2 \mathcal{N} T^3/45$, the ratio of the axion number density to the entropy at $t_w$ can be calculated as

$$\frac{n_a}{s} \approx 6.7 \times 10^6 \left[1 - \left(1 + \frac{\alpha}{\kappa}\right)^{-3/2}\right] \Delta \left(\frac{\bar{m}_a}{6 \times 10^{-6}\,\text{eV}}\right)^{-1} \left(\frac{f_a}{10^{12}\,\text{GeV}}\right)^{2.18},$$

using typical parameter values $\Gamma_a \approx 65$, $\nu \approx 0.40\zeta \alpha^{1/2}$ and $\zeta \approx 13$, Assuming number conservation after $t_w$ and using the entropy density $s_0 = 2809(T_0/2.7\text{K})^3 \text{cm}^{-3}$ and critical density $\rho_\text{crit} = 1.88 \times 10^{-29} h^2 \text{gcm}^{-3}$ at the present day, one can deduce that the axion loop contribution is

$$\Omega_{a,\ell} \approx 10.7 \left(\frac{\alpha}{\kappa}\right)^{3/2} \left[1 - \left(1 + \frac{\alpha}{\kappa}\right)^{-3/2}\right] h^{-2} \Delta \left(\frac{T_0}{2.7\text{K}}\right)^3 \left(\frac{f_a}{10^{12}\,\text{GeV}}\right)^{1.18}, \qquad (14)$$

where the Hubble's constant at the present day is $H_0 = 100h\,\text{km s}^{-1}\,\text{Mpc}^{-1}$, $0.35 < h < 1.0$,

## 4 Other contributions to the axion density

The contribution from long strings was roughly estimated in ref. [10]. The basis for this calculation was the radiation power per unit length for a typical sinusoidal perturbation, $dP/d\ell \approx \pi^3 f_a^2/16\gamma t$*, with the long string backreaction scale given by $\gamma \sim (\pi^2/8)[\ln(t/\delta)]^{-1}$. Assuming the radiative dominance of this smallest scale $\gamma t$ (as observed in ref. [17]), one can calculate the spectral density of axions from long strings

$$\frac{d\rho_a}{d\omega} \approx \frac{\pi^3 f_a^2 \zeta}{8\gamma \omega t^2}. \qquad (15)$$

---

* This also assumes that the energy lost by the long strings doesnot interfere with the scaling solution

Using similar methods to those used for loops

$$\Omega_{a,\infty} \approx 1.2 h^{-2} \Delta \left(\frac{T_0}{2.7\,\text{K}}\right)^3 \left(\frac{f_a}{10^{12}\,\text{GeV}}\right)^{1.18}, \qquad (16)$$

which, as before, is found to be roughly independent of the actual backreaction scale $\gamma$. The considerable uncertainty of (16) must be emphasised given its sensitivity to the amplitude of small-scale structure and the overall long string radiation spectrum.

A comparison of the two contributions (14) and (16) yields,

$$\frac{\Omega_{a,\ell}}{\Omega_{a,\infty}} \approx 8.9 \left(\frac{\alpha}{\kappa}\right)^{3/2} \left[1 - \left(1 + \frac{\alpha}{\kappa}\right)^{-3/2}\right]. \qquad (17)$$

Either of the contributions could be dominant for the expected parameter range, $0.1 < \alpha/\kappa < 1$ with equality at $\alpha/\kappa \approx 0.45$.

An order-of-magnitude estimate of the demise of the string/domain wall network [21] indicates that there is an additional contribution $\Omega_{a,\text{dw}} \sim (t_{\text{ann}}/\tilde{t})^{3/2}(f_a/10^{12}\,\text{GeV})$, where $t_{\text{ann}} \sim t_w$ is the time of wall annihilation. This 'domain wall' contribution is ultimately due to loops which are created at the time $\sim t_w$. Although the resulting loop density will be similar to (4), there is not the same accumulation from early times, so it is likely to be subdominant relative to (14). Both the long string and domain wall contributions will serve to strengthen the loop bound (14) on the axion; they are currently being studied in more detail [22].

## 5  Discussion and conclusions

The estimates for the loop and long string contributions to the axion density are summarized in fig. 1 as a function of the relative loop creation size $\alpha/\kappa$. This ratio expresses the key uncertainty arising from our inadequate understanding of long string radiative backreaction, $\gamma \lesssim \alpha \lesssim \kappa$. If we take the value implicitly assumed by most previous authors, $\alpha/\kappa \approx 1$, then requiring $\Omega_a < 1$ in the loop bound (14) we obtain a stringent constraint on the symmetry breaking scale,

$$f_a \lesssim 6.0 \times 10^{10}\,\text{GeV} \quad m_a \gtrsim 100\,\mu\text{eV}, \qquad h = 0.5, \qquad (18)$$

or $f_a \lesssim 1.9 \times 10^{11}\,\text{GeV}, m_a \gtrsim 31\,\mu\text{eV}$, for $h = 1.0$ (we have not included the parameter uncertainties of (6)). In this case, the axion is left a very narrow window which may be closed by the long string and domain wall contributions. However, as our recent investigations indicate [10], a parameter value as low as $\alpha/\kappa \sim 0.1$ may be more appropriate, that is, a regime in which the long string bound (16) is more important. This appears to provide an overall (uncertain) upper limit on the symmetry breaking scale,

$$f_a \lesssim 2.6 \times 10^{11}\,\text{GeV} \quad m_a \gtrsim 23\mu\text{eV}\,, \qquad h = 0.5\,, \tag{19}$$

or $f_a \lesssim 8.3 \times 10^{11}\,\text{GeV}\,, m_a \gtrsim 7\mu\text{eV}$, for $h = 1.0$. Note that the weakest string bound is stronger than the early homogeneous zero-momentum estimates, $f_a \lesssim 10^{12}\,\text{GeV}$ [3].

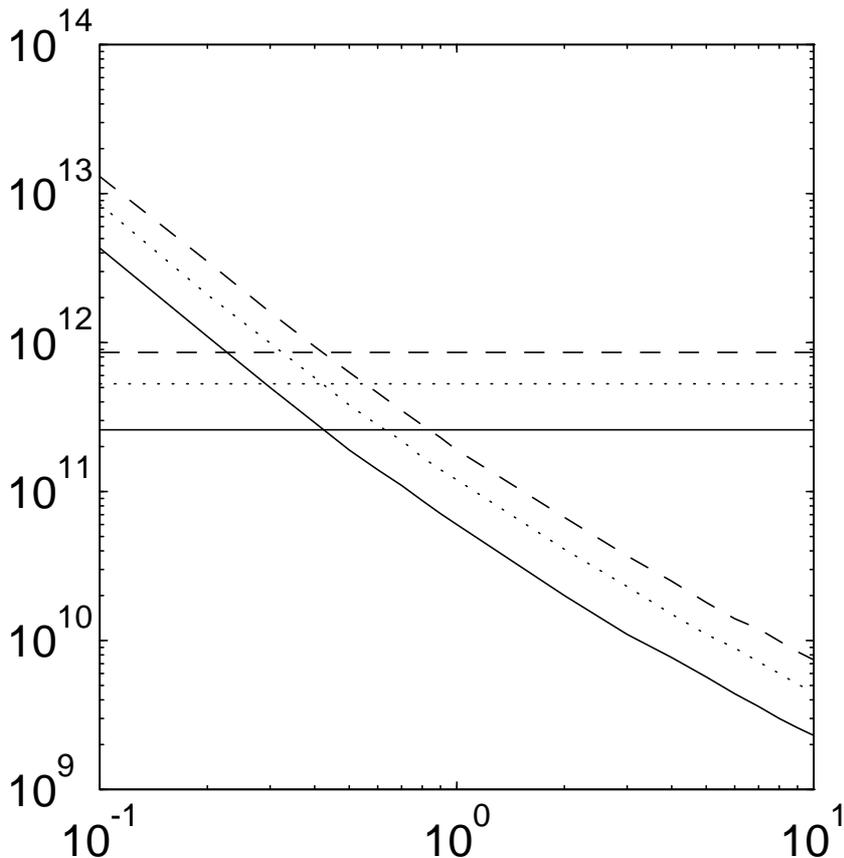

**Figure 1:** The loop bound (14) and the (constant) long string bound (16) on the symmetry breaking scale $f_a$ for various values of $\alpha/\kappa$. The solid line corresponds to $h = 0.5$, the dotted line to $h = 0.75$ and the dashed line to $h = 1.0$.

We have summarised a model for string evolution and radiative backreaction which should clarify and correct the methods by which axion string constraints are calculated. Considerable uncertainties remain and it is clearly a matter of some priority to resolve the axion string radiative backreaction issue [22]. Nevertheless, we conclude that axion emission by strings provides the key cosmological constraint on $f_a$ and $m_a$ in the standard scenario. A window remains for experimental searches for the axion above the astrophysical bound $f_{\rm a} \gtrsim 10^9$–$10^{10}$ GeV, but it is tightly constrained.

## Acknowledgements

We are grateful for helpful discussions with Alex Vilenkin, David Lyth, Georg Raffelt, Scott Thomas and Michael Turner. The loop bound (14) is a detailed calculation of a simple estimate in ref. [16].

## References

1. Peccei, R.D., & Quinn, H.R. [1977], *Phys. Rev. Lett.* **38**, 1440; *Phys. Rev.* **D16**, 1791.

2. Weinberg, S. [1978], *Phys. Rev. Lett.* **40**, 223. Wilczek, F. [1978], *Phys. Rev. Lett.* **40**, 279.

3. Abbott, L.F., & Sikivie, P. [1983], *Phys. Lett.* **120B**, 133. Preskill, J., Wise, M.B., & Wilczek, F. [1983], *Phys. Lett.* **120B**, 127.

4. Raffelt G. [1990], *Phys. Rep.* **198**, 1. Turner M. [1990], *Phys. Rep.* **197**, 678.

5. Vilenkin, A., & Everett, A.E. [1982], *Phys. Rev. Lett.* **48**, 1867.

6. Linde, A.D. [1991], *Phys. Rev. Lett.* **259**, 38.

7. Davis, R.L. [1985], *Phys. Rev.* **D32**, 3172. Davis, R.L. [1986], *Phys. Lett.* **180B**, 225.

8. Harari, D., & Sikivie, P. [1987], *Phys. Lett.* **195B**, 361. Hagmann, C., & Sikivie, P. [1991], *Nucl. Phys.* **B363**, 247.

9. Davis, R.L., & Shellard, E.P.S. [1989], *Nucl. Phys.* **B324**, 167.

10. Battye, R.A., & Shellard, E.P.S. [1994], to appear in *Nucl. Phys.* **B**.

11. Bennett, D.P., & Bouchet, F.R. [1990], *Phys. Rev.* **D41**, 2408.

12. Allen, B., & Shellard, E.P.S. [1990], *Phys. Rev. Lett.* **64**, 119. Shellard, E.P.S., & Allen, B. [1990], in *Formation and Evolution of Cosmic Strings*, Gibbons, G.W., Hawking, S.W., & Vachaspati, T., eds. (Cambridge University Press).

13. Vilenkin, A., & Vachaspati, T. [1987], *Phys. Rev.* **D35**, 1138.

14. Sakellariadou, M. [1991], *Phys. Rev.* **D44**, 3767.

15. Shellard, E.P.S. [1986], in Proceedings of the 26th Liege International Astrophysical Colloquium, *The Origin and Early History of the Universe*, Demaret, J., ed. (University de Liege). Shellard, E.P.S. [1990], in *Formation and Evolution of Cosmic Strings*, Gibbons, G.W., Hawking, S.W., & Vachaspati, V., eds. (Cambridge University Press). Thomas S. [1993], PhD Thesis, Univ. Texas.

16. Vilenkin, A. & Shellard, E.P.S. [1994], *Cosmic strings and other topological defects* (Cambridge University Press (*in press*)).

17. Allen, B., & Shellard, E.P.S. [1992], *Phys. Rev.* **D45**, 1898.

18. Sikivie, P. [1982], *Phys. Rev. Lett.* **48**, 1156.

19. Turner M. [1986], *Phys. Rev.* **D33**, 889.

20. Ryden, B.S., Press, W.H., & Spergel, D.N. [1990], *Ap. J.* **357**, 293.

21. Lyth, D.H. [1992], *Phys. Lett.* **275B**, 279.

22. Battye, R.A., & Shellard, E.P.S. [1994], in preparation